\documentclass[twocolumn,preprintnumbers,amsmath,amssymb]{revtex4}
\usepackage{color}
\usepackage{bm, graphicx, amsmath}
\usepackage{hyperref}
\def\mathLarge#1{\mbox{\Large $#1$}}
\def\mathlarge#1{\mbox{\large $#1$}}

\begin{document}

\title{Simulating Quantum Algorithms Using Fidelity and Coherence Time as Principle Models for Error}
\author{Daniel Koch$^{1}$, Avery Torrance$^{2}$, David Kinghorn$^{2}$, Saahil Patel$^{1}$, Laura Wessing$^{1}$, Paul M. Alsing$^{1}$}
\affiliation{$^{1}$Air Force Research Lab, Information Directorate, Rome, NY}
\affiliation{$^{2}$Rochester Institute of Technology, Rochester, NY}

\begin{abstract}

As various quantum computing technologies continue to compete for quantum supremacy, several parameters have emerged as benchmarks for the quality of qubits. These include fidelity, coherence times, connectivity, and a few others. In this paper, we aim to study the importance of these parameters and their impact on quantum algorithms. We propose a realistic connectivity geometry and form quantum circuits for the Bernstein-Vazirani, QFT, and Grover Algorithms based on the limitations of the chosen geometry. We then simulate these algorithms using error models to study the impact of gate fidelity and coherence times on success of the algorithms. We report on the findings of our simulations and note the various benchmarking values which produce reliably successful results.

\end{abstract}

\maketitle

%
%
\section{Introduction}
%

In the years to come, the race for bigger and better quantum computers will yield a new plethora of NISQ (noisy intermediate-scale quantum) devices, nearing the milestone of 100 qubits. With major commercial players such as IBM, Google, and several others \cite{ibm,google,rigetti,microsoft} competing to drive the technological limitations of these machines, the need for standardizing parameters by which users can compare qubit qualities has risen.  Analogous to the way in which classical computers are categorized by meaningful criteria: CPU, GPU, RAM, etc., the field of quantum computing has naturally gravitated towards the quantities: fidelity \cite{nielsen}, coherence times \cite{T1_T2,T2_1,T2_2}, and connectivity. In this paper, we set out to simulate noisy quantum systems using these parameters and draw predictions about their influence on future algorithm success.

Popularized by the phrase 'Quantum Supremacy', the ultimate goal for quantum computing is the realization of computations through qubits which are either faster than or intractable by classical means.  However, in order to progress quantum computing to these milestones, the study of these quantum systems through classical methodologies is needed.  While some aspects of quantum systems can be simulated through numerical techniques \cite{jozsa,gao}, often times the full simulation of larger systems is simply intractable.  As a result, approximate noise models are used when applicable, which in turn has sparked a growing number of tools for simulating various quantum systems \cite{markov,radtke}, including some by the commercial players.

Because noisy qubits are an unavoidable reality for the coming NISQ era years, simulations of these error-prone systems seek to advance the field of quantum computing through one of two important avenues.  First, in order to create better qubits one needs to understand the fundamental sources of noise which exist due to qubit interactions with their environment \cite{carlo,karmen,crow}.  Simulations of this nature seek to improve qubit technologies by modeling the underlying physics which contribute to qubit errors, such as decoherence and depolarization, making up the larger contribution of studies to this field.

As a second important motivation for simulating quantum systems with noise, the surge of available quantum computers over the past several years has many investigating the potential near and future uses of these devices.  While there are several already known theoretical speedups for these quantum computers \cite{G,shor}, the realization of these algorithms on current hardware \cite{linke,coles}, or simplified versions thereof, demonstrates that the impact of noise on algorithm success \cite{tannu,burnett,koch} cannot be ignored.  Consequently, many have recognized the need for co-development of quantum algorithms with noisy qubits \cite{kapit,murphy}, developing new protocols and techniques for improving algorithm success by minimizing the impact of errors.

In this paper, our primary motivation for simulating quantum systems is to investigate the degree to which errors related to gate fidelities, energy relaxation, and measurement collapse inhibit the success of quantum algorithms.  The specifics for how each noise model is simulated are described in the coming sections, with a consistent theme whereby each error type is governed by a single parameter.  We then show through our simulations the different benchmarking breakpoints for each parameter, or combinations thereof, in order to achieve various levels of algorithmic success. 

\subsection{Layout}
%

In section 2 we outline the specifics of the quantum systems we simulate, namely connectivity constraints and circuit diagrams for the algorithms studied \cite{G,BV,QFT}. In section 3, we outline our model for coherent noisy gates. We provide full mathematical descriptions for our implementation of these noisy gates and their relation to the parameter $f$. Section 4 contains the results of simulating the various quantum algorithms using our noisy gates model. In section 5, we outline our methodology for implementing decoherent errors into the simulations, whereby qubits can probabilistically collapse according analogous to energy relaxation and partial measurements. Section 6 discusses the results of these errors and their impact on algorithm success when they are the only source of error. In section 7 we combine both of the previously studied error models, showcasing how each algorithm performs under these constraints. Section 8 is a concluding summary of the results found throughout the paper as well as a discussion of potential future work.

%
%
\section{Qubit Geometries and Circuits}
%

\subsection{Connectivity}
%

As mentioned in the introduction, everything that will go into the simulations throughout this paper is constructed with current hardware metrics and limitations in mind. Thus, we will begin by discussing a limitation on current quantum computers that indirectly affects quantum algorithm success, qubit connectivity. Different quantum computing technologies offer various qubit connectivity, some better than others. In this paper we will be basing our simulations with superconducting qubits in mind, which typically have qubit connections around the 3-5 range.

We will propose a limited qubit geometry here that is on par with current hardware, shown below in figure \ref{Fig:QubitGeometry}. Consequently, for the quantum algorithms in this study, we adapt the idealized versions of these algorithms to run on this chosen geometry. Doing so requires the use of additional gates in order to carry out 2-qubit operations between qubits that do not share a direct connection.

\begin{figure}[h] 
\centering
\includegraphics[scale=0.1]{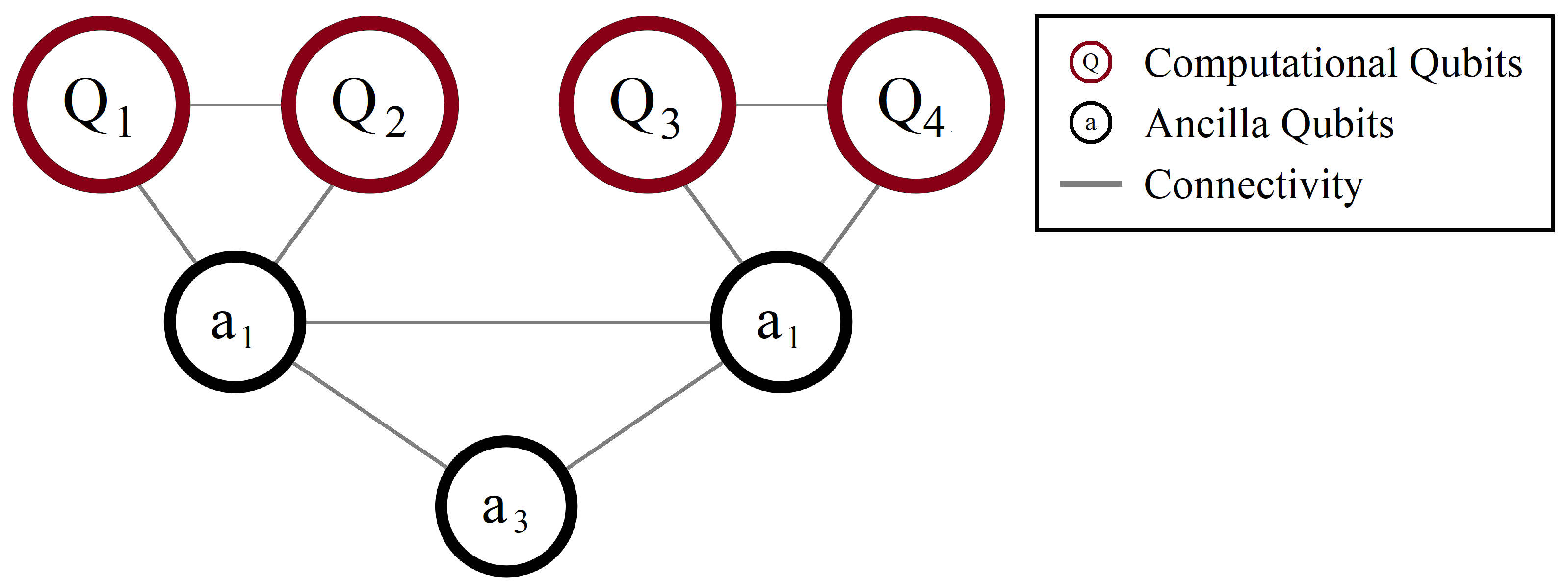}
\caption{Top row of qubits marked by ``Q'' (dark red): computational qubits. Lower qubits marked by ``a'' (black): ancilla qubits. The computational qubits represent the main quantum system where each algorithm will take place, while the role of the supporting ancilla qubits is to act as intermediates for multi-qubit gate operations between computational qubits which do not have a direct connection.}
\label{Fig:QubitGeometry}
\end{figure}

The motivation for the geometry shown in figure \ref{Fig:QubitGeometry} is twofold: 1) The connections shown can be mapped to several current hardware designs (for example, IBM's 20-qubit chip ``Tokyo''), requiring qubits only have at most four nearest neighbor connections. 2) This geometry is scalable up to any size for producing $2^N$ computational qubits, requiring $2^{N}-1$ ancilla qubits. Most importantly, higher orders of $N$ do not require more connectivity, only more qubits. The computational qubits require a connectivity of 2 (top row of qubits in figure \ref{Fig:QubitGeometry}), while the ancilla qubits require 4 (except for the very bottom-most ancilla qubit). For another example, a geometry for $N=3$ is shown below in figure \ref{Fig:QubitGeometry2}:

\begin{figure}[h] 
\centering
\includegraphics[scale=0.08]{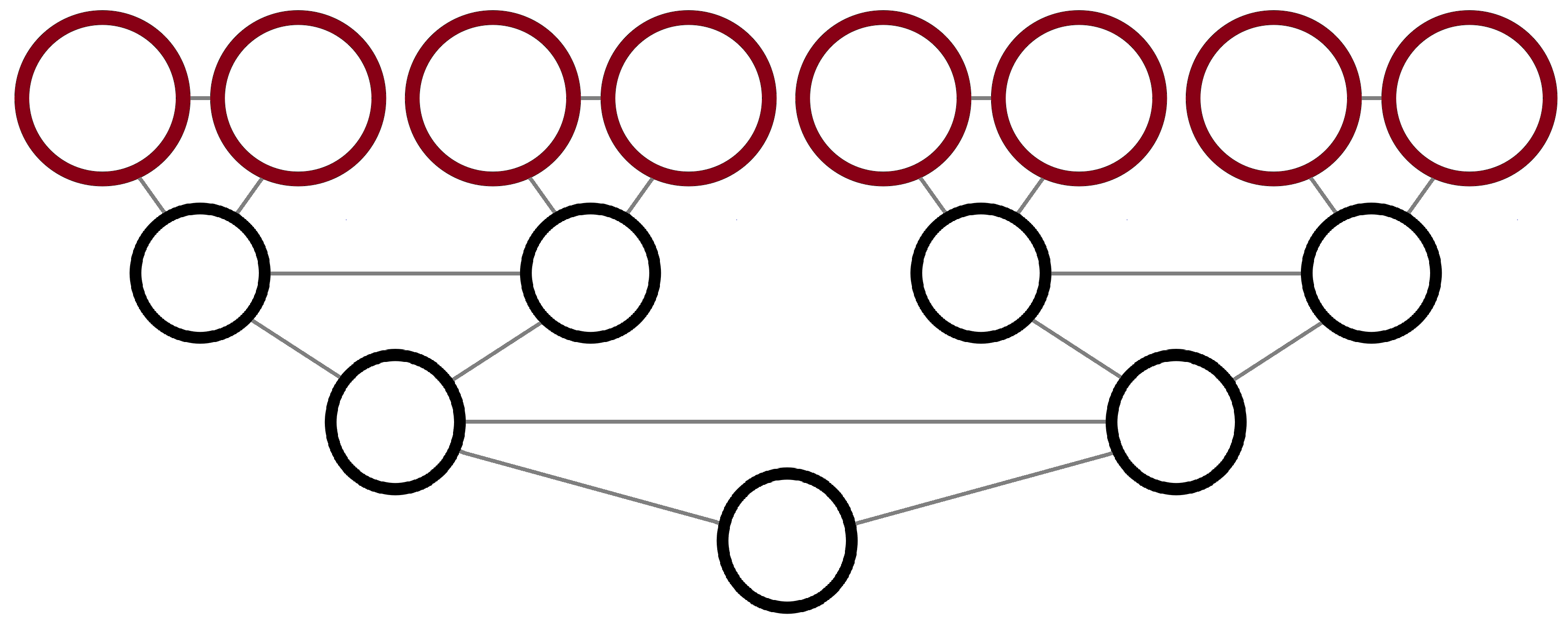}
\caption{$N=3$ qubit geometry. The top layer consists of 2$^3$ computational qubits, requiring 7 ancilla qubits.}
\label{Fig:QubitGeometry2}
\end{figure}

The tradeoff for this scalable geometry comes in two forms: 1) 2 or 3-qubit gate operations between distant computational qubits require increasingly more intermediate quantum gates, resulting in potentially more errors from imperfect gate operations and overall longer quantum circuits. 2) Working with $2^N$ computational qubits requires a total qubit geometry of nearly double size, making the overall algorithms twice as sensitive to coherence errors.

Using figure \ref{Fig:QubitGeometry2} as an example, one can see that the number of connections separating some of the computational qubits is as high as five. This means that a 2-qubit gate between such qubits would require five times as many operations (often more), drastically increasing the chance of errors impacting the algorithm. Simultaneously, these longer operations require more time and qubits to achieve, opening up more possibilities for both the computational and ancilla qubits to decohere. Nevertheless, we have chosen these qubit geometries, with all their faults, such that we may study the way in which these connectivity restraints impact quantum algorithm success.

\subsection{Algorithm Circuits}
%

Because of limited connectivity, we must adapt the idealized versions of each quantum algorithm to fit our particular geometry choice. In general, these adapted versions follow all of the same steps as the idealized algorithms, but require additional control gates to and from the ancilla qubits. The quantum circuits for the Berstein-Vazirani, Grover, and QFT algorithms \cite{BV,G,QFT} are shown below in figures \ref{Fig:BVCircuit} - \ref{Fig:GroverCircuit}, for the case of an $N=2$ geometry ($2^2$ computational qubits).

\begin{figure}[h] 
\centering
\includegraphics[scale=0.5]{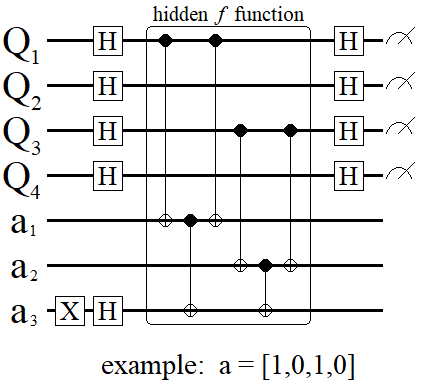}
\caption{Quantum circuit for the Bernstein-Vazirani Algorithm, shown for the case where $a = [1,0,1,0]$ ($a$ is the hidden bit-string). Note that some of the CNOT gates in this circuit can be parallelized in order to shorten the circuit depth, but our simulations do not do so.}
\label{Fig:BVCircuit}
\end{figure}

\begin{figure}[h] 
\centering
\includegraphics[scale=0.5]{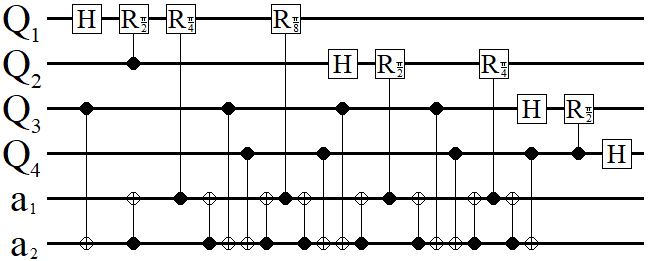}
\caption{Quantum Circuit for a Quantum Fourier Transformation. When studying this circuit throughout the paper, only the gate operations shown here are subject to fidelity and coherence errors. Additionally, we do not include the standard $\textbf{SWAP}$ gates at the end of the circuit.}
\label{Fig:QFTCircuit}
\end{figure}

\begin{figure}[h] 
\centering
\includegraphics[scale=0.5]{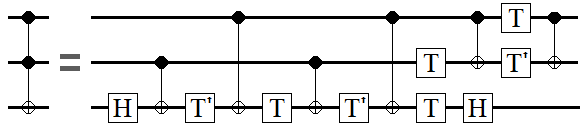}
\caption{Decomposition of the CCNOT gate into 1 and 2-qubit operations. This circuit is used in place of all CCNOT gates found in other circuits.}
\label{Fig:CCNOTCircuit}
\end{figure}

\begin{figure}[h] 
\centering
\includegraphics[scale=0.5]{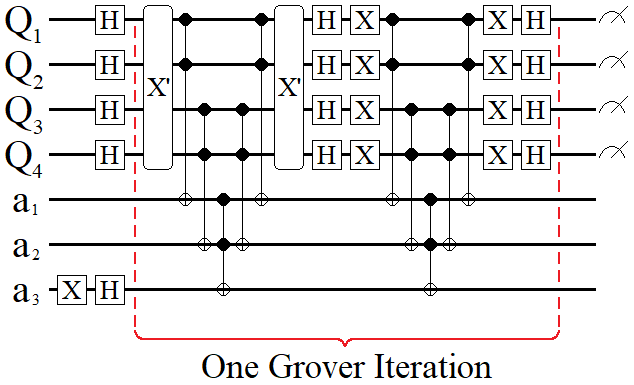}
\caption{Quantum Circuit for the Grover Algorithm. The operator $\textbf{X'}$ represents the application of $\textbf{X}$ gates that correspond to the desired state (for example, searching for the state $|0101\rangle$ would require $\textbf{X'} = \textbf{X}_2 \otimes \textbf{X}_4$). We note that there is one difference between this circuit diagram and the one run in the simulations, and that is that several of the CCNOTS are parallelized where applicable in order to minimize the total circuit time.}
\label{Fig:GroverCircuit}
\end{figure}

All of the circuits shown are the exact instructions used in our simulations. Obeying the geometry laid out in figure \ref{Fig:QubitGeometry}, in conjunction with the gates \textbf{X}, \textbf{H}, \textbf{T}, \textbf{R}$_{\phi}$, and \textbf{CNOT}, the circuits presented here are all in principle realizable on any available quantum computing hardware that can support the required connectivity. Thus, all of the simulation results obtained in the following sections are comparable with potential experimental results.

%
%
\section{Coherent Noisy Gates}
%

When evaluating different quantum computing technologies in terms of quality, often times gate fidelity is the first metric people gravitate towards. Justifiably so, quantum algorithms require precise gate operations in order to maximize the advantages that superposition states allow for. Thus, identifying ``how good'' a quantum computer's gates are is a natural first benchmark. The parameter fidelity ($f$) is most often used to classify this metric, defined in several closely related ways \cite{nielsen} depending on the type of quantum system being studied, but generally always interpreted as the ``closeness'' between two quantum states. In this study we will associate the parameter $f$ with each quantum gate, denoting how close a particular gate operation transforms a qubit(s) to the intended final state:

\begin{eqnarray}
U | \Psi \rangle &=& | \Phi \rangle \nonumber \\
\tilde{U} | \Psi \rangle &=& | \phi \rangle \nonumber \\
f &=& | \langle \Phi | \phi \rangle | ^2
\label{Eqn:fidelity}
\end{eqnarray}

Equation \ref{Eqn:fidelity} above shows the definition of fidelity between the two pure states $|\Phi \rangle$ and $|\phi \rangle$, where $U$ is some theoretical gate operation and $\tilde{U}$ represents an imperfect version of the same gate. $\tilde{U}$ carries an error with it, achieving some final state differing from $| \Phi \rangle$, which we will define as a coherent error for this paper. Specifically, a coherent error is one where an imperfect gate operation can be modeled by a unitary operator. These coherent errors result in pure states that are skewed in some way, such that their overlap with the intended output state defines the gate's fidelity.

Supposing we would like to determine $f$ for some unitary gate $U$ experimentally, one simple way is to apply $U U^{\dagger}$ to a qubit(s) and then make a measurement. In principle, applying such an operation should always return a qubit back to its original state, $|0\rangle$ in most cases. However, experimentally one may occasionally find the state $|1\rangle$, implying that the operation $U U^{\dagger}$ did not transform the qubit's state as intended (assuming one can rule out other sources of error). Repeating this process many times, one can determine an average fidelity $\langle f \rangle$, which is the value most often reported. It is important to note that on any given individual application of some $U$, we cannot say for certain if the operation was successful or not, thus we must most often discuss fidelities in terms of averages.

\subsection{Coherent Amplitude Error}
%

In the experiment just described, there are several contributing factors as to why one might measure the $|1\rangle$ state (when expecting to find $|0\rangle$). The issue is that it is very difficult to pin down quantum errors to a single source. The interactions that a qubit has with gates, other qubits, and the environment are all very delicate and intertwined. Thus, the aim of our study here is to simulate each quantum algorithm using models that isolated isolate each source of error.

To begin, our first error model focuses solely on the quantity of fidelity and its relationship to imperfect gate operations. Specifically, we will study a model for imperfect gates whereby the error occurs on the amplitudes of the output state. Let us define the parameter $\epsilon$, which denotes the amount of amplitude that ends up on an unintended component of the final state. For 1-qubit gates, we can express the effect of our coherent error gates, $\tilde{\textbf{U}}_{\epsilon}$, as follows:

\begin{equation}
\tilde{\textbf{U}}_{\epsilon} |\Psi \rangle = \epsilon | \Phi_{\perp} \rangle + \sqrt{1 - \epsilon ^2}|\Phi ' \rangle
\label{Eqn:ImperfectU}
\end{equation}

Here, $| \Phi ' \rangle$ represents the intended output state one would expect from $\textbf{U}$, and $| \Phi_{\perp} \rangle$ is the state orthogonal to $|\Phi ' \rangle$. Specifically, $| \Phi_{\perp} \rangle$ is determined by replacing $| \Psi \rangle$ with a bit-flip error (all $|0\rangle$ and $|1\rangle$ components switch) and applying a perfect $\textbf{U}$ operation. Below is a graphical representation of such an error transformation, using the $\textbf{X}$ gate as an example:

\begin{figure}[h] 
\centering
\includegraphics[scale=0.4]{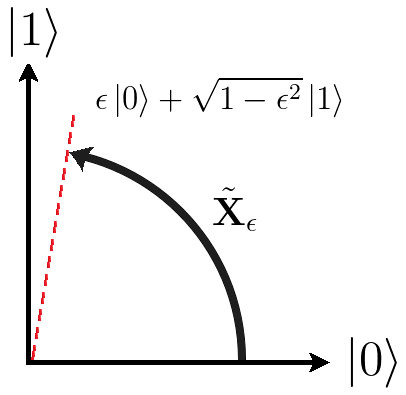}
\caption{Graphical representation of the $\tilde{\textbf{X}}_{\epsilon}$ operation on the state $|0\rangle$. Here, $ | \Psi \rangle = | 0 \rangle$, $| \Phi ' \rangle = | 1\rangle$, and $| \Phi_{\perp} \rangle = \textbf{X} |1\rangle = |0\rangle$. As shown in the final state, a small amplitude $\epsilon$ resides on the $|0\rangle$ component. }
\label{Fig:XeTransformation}
\end{figure}

Figure \ref{Fig:XeTransformation} shows an example of a coherent amplitude error, whereby the $\tilde{\textbf{X}}_{\epsilon}$ transformation results in a final state that is displaced from the ideal final state by an angle of $sin^{-1}(\epsilon)$. Using the definition of fidelity from equation \ref{Eqn:fidelity} in conjunction with $\tilde{\textbf{U}}_{\epsilon}$ from equation \ref{Eqn:ImperfectU}, we get the following relation for gate fidelity: $f = 1 - \epsilon^2$. Thus, when $\epsilon = 0$, we recover the perfect $\textbf{U}$ gate and obtain a fidelity of $1$. Also note that both $\epsilon$ and -$\epsilon$ result in the same fidelity, a result that will be more important later on.

\begin{figure}[h] 
\centering
\includegraphics[scale=0.5]{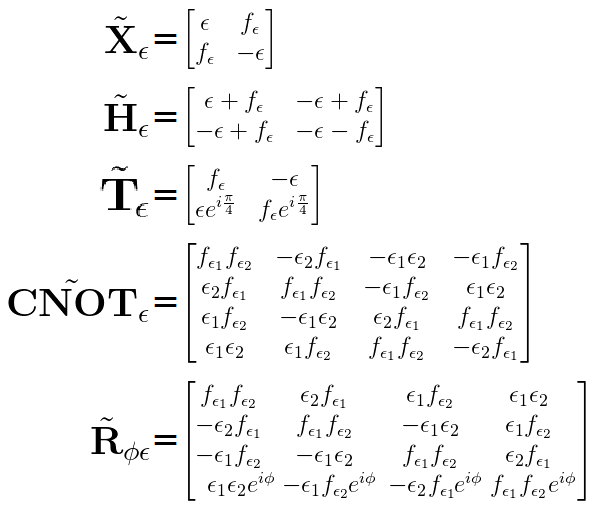}
\caption{Matrix representations of each 1 and 2-qubit gate with coherent amplitude errors. For the 2-qubit gates there are two sources of possible errors, $\epsilon_1$ and $\epsilon_2$, representing the potential for incorrect amplitudes on both the target and control qubits ($\epsilon_1$ for the control, $\epsilon_2$ for the target). In all matrices we have used the following shorthand:$f_{\epsilon} \equiv \sqrt{1-\epsilon^2}$.}
\label{Fig:AmpErrorGates}
\end{figure}

We can extend the formalism outlined in equation \ref{Eqn:ImperfectU} to create coherent error versions of all the gates in figures \ref{Fig:BVCircuit} - \ref{Fig:GroverCircuit}. Each gate follows the same guiding principle, whereby the parameter $\epsilon$ represents the amplitude error residing on the state orthogonal to the intended final state. The matrix forms for all of the coherent amplitude error gates are given in figure \ref{Fig:AmpErrorGates}

In figure \ref{Fig:AmpErrorGates}, certain $\epsilon$ values are given negative signs in order to keep each matrix unitary. However, these negative signs do not affect the magnitude of the errors in any way, only differing in the direction in which the error occurs (see figure \ref{Fig:XeTransformation}). By allowing for equal likeliness of both positive and negative values for $\epsilon$, these negative signs have no impact on the gates. Note that each of these error matrices returns to their respective theoretical versions for the case where $\epsilon = 0$, becoming perfect gates for fidelity values of $1$. Using the error gates shown in figure \ref{Fig:AmpErrorGates}, we get the following fidelities:

\begin{eqnarray}
f_{\textrm{1-qubit}} &=& 1 - \epsilon^2 \label{Eqn:1QubitFid} \\
f_{\textrm{2-qubit}} &=& (1 - \epsilon_1^2)(1 - \epsilon_2^2) \label{Eqn:2QubitFid}
\end{eqnarray}

\subsection{Average Fidelity and Sampling}
%
	
Using the error gates defined in the previous section, we must now specify exactly how we are implementing these $\epsilon$'s into our quantum circuits. Ideally, one would choose $\epsilon$ values associated with the fidelities of each gate, mimicking each gate's tendencies to contribute to the overall error in the system. However, we cannot know the exact amplitude errors for any real system, so we are instead forced to work with averages. Additionally, we should assume that repeated uses of the same gate do not result in the exact same amplitude error, so we cannot assign a single $\epsilon$ for each gate.

Because average fidelities are the standard quantity often reported for quantum computers, we will incorporate them into our model here. In particular, our simulations will assign each individual application of a gate a randomly chosen fidelity, some better than others, ultimately averaging out to $\langle f \rangle$ for a given gate. We incorporate this randomness through the parameter $\epsilon$, whereby the simulation selects random $\epsilon$ values from some probability distribution, P($\epsilon$). The only requirement on this probability distribution is that it must give rise to the expected macroscopic value for $\langle f \rangle$ through random sampling. Rewriting \ref{Eqn:1QubitFid} and \ref{Eqn:2QubitFid} in terms of averages, we get equations which set the constraints on choosing probability distributions:

\begin{eqnarray}
\langle f_{\textrm{1-qubit}} \rangle &=& 1 - \langle \epsilon^2 \rangle \label{Eqn:1QubitFidAvg} \\
\langle f_{\textrm{2-qubit}} \rangle &=& 1 - \langle \epsilon_1^2 \rangle - \langle \epsilon_2^2 \rangle + \langle \epsilon_1^2 \epsilon_2^2 \rangle \label{Eqn:2QubitFidAvg}
\end{eqnarray}

So long as one samples $\epsilon$'s from a probability distribution that satisfies equations \ref{Eqn:1QubitFidAvg} and \ref{Eqn:2QubitFidAvg}, the coherent error gates will reflect any value chosen for $\langle f \rangle$. Ideally then, we would like to sample from a probability distribution that accurately reflects the underlying nature of each gate's error tendencies. However, it is difficult to say what the nature of such a distribution might be, especially when considering the same quantum gate achieved through various technologies.

%
%
\section{Fidelity Gates Analysis}
%

\subsection{Role of Probability Distributions}
%

In principle, although two P($\epsilon$) distributions may result in the same $\langle f \rangle$, the way in which they represent errors could impact a quantum algorithm differently. So then, in order to understand the role that an underlying P($\epsilon$) distribution may have, we will study two probability distributions that possess distinct differences. Each distribution satisfies equations \ref{Eqn:1QubitFidAvg} and \ref{Eqn:2QubitFidAvg}, as well as $\langle \epsilon \rangle = 0$, implying that the average $\epsilon$ value has no bias (the errors are symmetric in the way they deviate from the intended final state). Both distributions are modeled as gaussians, but the major distinction between them lies in where the most probable $\epsilon$'s occur.

\begin{eqnarray}
\textrm{P}_1(\epsilon) &=& \frac{1}{ \sqrt{2 \sigma_1^2} } e^{- \mathlarge{ \frac{{\epsilon}^2}{2 \sigma_1^2}} } \label{Eqn:P1(e)} \\
\textrm{P}_2(\epsilon) &=& \frac{1}{2 \sqrt{2 \sigma_2^2} } \Huge{(}e^{- \mathlarge{ \frac{(\epsilon-\bar{\epsilon} )^2}{2 \sigma_2^2}} } +e^{- \mathlarge{ \frac{(\epsilon+\bar{\epsilon} )^2}{2 \sigma_2^2}} } \Huge{)}
\label{Eqn:P2(e)}
\end{eqnarray}

\begin{figure}[h] 
\centering
\includegraphics[scale=0.42]{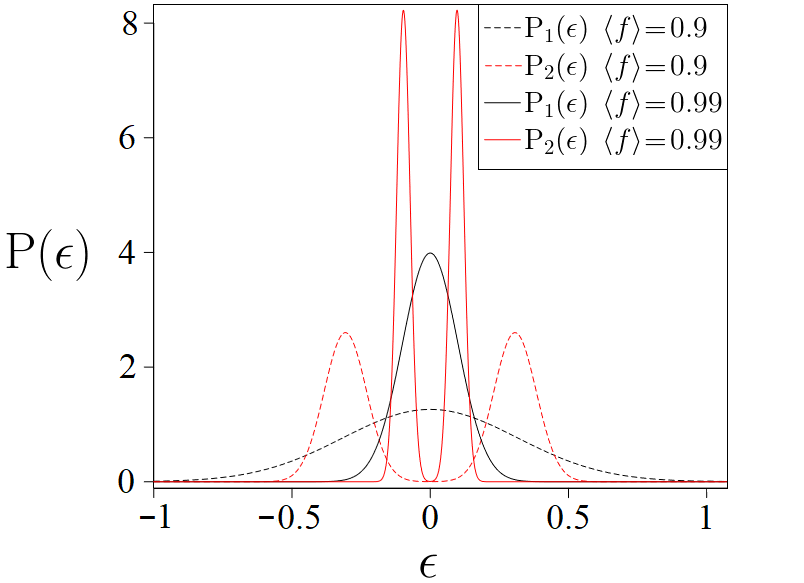}
\caption{Plots for P$_1$($\epsilon$) and P$_2$($\epsilon$) with fidelity values of 0.9 and 0.99. Both distributions extend out to the infinity, but for our simulations we only use values where $|\epsilon| < 1$.}
\label{Fig:PDistributions}
\end{figure}

P$_1$($\epsilon$) corresponds to an underlying error model where $\epsilon = 0$ is the most probable value, representing the situation where gate operations are most often close to a fidelity of 1, but the chance for large $\epsilon$ errors are still non-negligible. Conversely, P$_2$($\epsilon$) reflects the case where the most probable $\epsilon$ values are centered around the peaks $\pm \bar{\epsilon}$, which become closer to $\epsilon = 0$ as fidelity approaches 1. This distribution represents the situation where large $\epsilon$ errors are much less common, but so too are values close to $\epsilon = 0$. Plotted in figure \ref{Fig:PDistributions} below are P$_1$($\epsilon$) and P$_2$($\epsilon$) for various fidelity values.

Looking at figure $\ref{Fig:PDistributions}$, the important difference to note is where the majority of each probability distribution is concentrated. For completeness, the values for $\sigma_1$, $\sigma_2$, and $\bar{\epsilon}$ for equations $\ref{Eqn:P1(e)}$ and $\ref{Eqn:P2(e)}$ are given below.

\begin{eqnarray}
\sigma_1 &=& \sqrt{ 1 - \langle f \rangle } \\
\sigma_2 &=& \bar{\epsilon}/4 \\
\bar{\epsilon} &=& \sqrt{ \frac{16}{17}( 1 - \langle f \rangle ) }
\end{eqnarray}

Note that while P$_1$($\epsilon$) and P$_2$($\epsilon$) are true probability distributions, in our simulations we do not allow for values of $|\epsilon| > 1$. This constraint on $\epsilon$ is required by our error gate models in order to stay unitary. Incidents of $|\epsilon| > 1$ in our random sampling are thrown out and a new $\epsilon$ value is simulated. For reference, at the lower bound of our simulations of $\langle f \rangle = 0.9$, the probabilities of picking an $| \epsilon | > 1$ are $10^{-3}$ and $10^{-6}$ for P$_1$($\epsilon$) and P$_2$($\epsilon$) respectively. By $\langle f \rangle = 0.99$, these probabilities become smaller than $10^{-10}$.

For the 2-qubit gates, we again use the probability distributions \ref{Eqn:P1(e)} and \ref{Eqn:P2(e)}, where we will assume the errors on each qubit are independent but still determined solely by a single average fidelity (equation \ref{Eqn:2QubitFidAvg}). Specifically, we have:

\begin{equation}
P_n(\epsilon_1,\epsilon_2) = P_n(\epsilon_1) P_n(\epsilon_2) \hspace{.6cm} n \in [1,2]
\label{Eqn:P(e1,e2)}
\end{equation}

Equation \ref{Eqn:P(e1,e2)} reflects that in our model the error for each qubit is independent of the other, but both contribute to the overall fidelity of the gate. The subscript $n$ in the equation refers to the two probability distributions $\ref{Eqn:P1(e)}$ and $\ref{Eqn:P2(e)}$. Substituting P$_1$ and P$_2$ into this equation and integrating gives us the following average fidelities for P$_1$($\epsilon_1$,$\epsilon_2$) and P$_2$($\epsilon_1$,$\epsilon_2$):

\begin{eqnarray}
P_1(\epsilon_1,\epsilon_2): \hspace{.5cm} \langle f \rangle &=& \big{(} 1 - \sigma_1^2 \big{)} \big{(} 1 - \sigma_2^2 \big{)} \label{Eqn:P1(e1e2)} \\
P_2(\epsilon_1,\epsilon_2): \hspace{.5cm}\langle f \rangle &=& \big{(} 1 - \frac{17}{16} \bar{\epsilon}_1^2 \big{)} \big{(} 1 - \frac{17}{16} \bar{\epsilon}_2^2 \big{)} \label{Eqn:P2(e1e2)}
\end{eqnarray}

Here, the subscripts $_1$ and $_2$ on the $\sigma$'s and $\bar{\epsilon}$'s refer to qubits 1 and 2 (1 for the control qubit, 2 for the target). Equations \ref{Eqn:P1(e1e2)} and \ref{Eqn:P2(e1e2)} are general, allowing for different $\sigma$ and $\bar{\epsilon}$ values for each of the qubits, which could be motivated experimentally. Here, we will assume that the inherent probabilities for error on the control and target qubits are equal. Setting these quantities to be equal results in the values given below.

\begin{eqnarray}
\sigma_1 = \sigma_2 &=& \sqrt{ 1 - \langle f \rangle^{\frac{1}{2}} } \label{Eqn:sigmas} \\
\bar{\epsilon}_1 = \bar{\epsilon}_2 &=& \sqrt{ \frac{16}{17} \big{(} 1 - \langle f \rangle^{\frac{1}{2}} \big{)} } \label{Eqn:epsilons}
\end{eqnarray}

\begin{figure}[h] 
\centering
\includegraphics[scale=0.4]{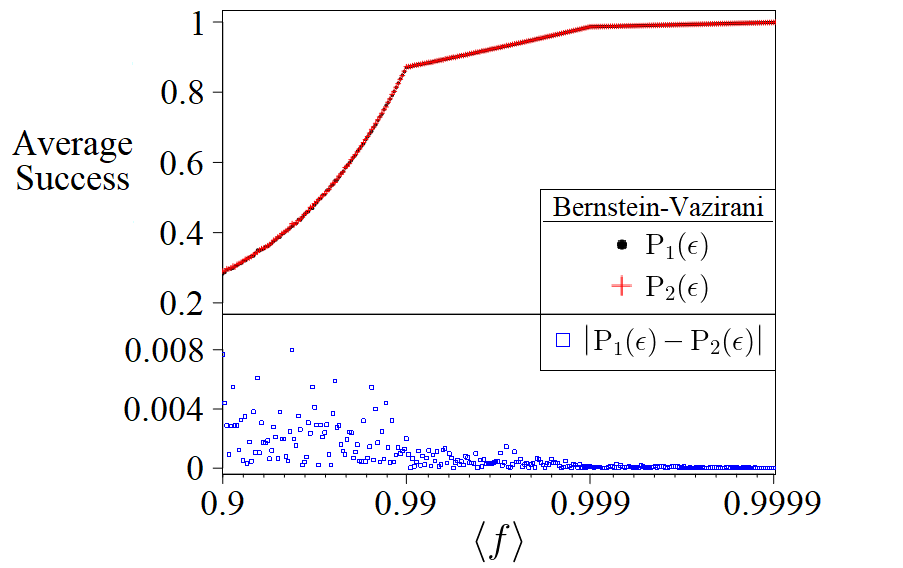}
\caption{ (top) Average success of the Bernstein-Vazirani Algorithm for $\langle f \rangle$ values ranging from 0.9 to 0.9999. Each data point represents the average from 10000 simulations. (bottom) The difference in success between P$_1$($\epsilon$) and P$_2$($\epsilon$) for each data point in the top plot.}
\label{Fig:P1vP2}
\end{figure}

By substituting the values in $\ref{Eqn:sigmas}$ and $\ref{Eqn:epsilons}$ into P$_1$($\epsilon_1$,$\epsilon_2$) and P$_2$($\epsilon_1$,$\epsilon_2$), we now how our complete formalism for simulating coherent noisy gates with the two different underlying probability distributions for $\epsilon$. Plotted in figure \ref{Fig:P1vP2} are the findings of our simulations for the Bernstein-Vazirani algorithm. The figure shows the influence of P$_1$($\epsilon$) and P$_2$($\epsilon$) for various fidelity values, as well as the difference between them.

As figure \ref{Fig:P1vP2} suggests, the two differing underlying probability distributions seem to have no overall impact on the success of the algorithm (we discuss our metric for determining algorithm success at the start of the next section). In the region where $0.9 < \langle f \rangle < 0.99$, the difference between average successes is at most $0.008$ (0.8\% success probability). These differences become negligible by the point $\langle f \rangle = 0.99$ and beyond. This result suggests that our fidelity model is strongly governed by $\langle f \rangle$, and not any particular P$_i$($\epsilon$).

As a final note, the results from figure $\ref{Fig:P1vP2}$ seem to suggest that any probability distribution that satisfies the condition $\langle \epsilon \rangle = 0$ will lead to the same average success. However, working with an underlying P($\epsilon$) that does not meet this condition ($\epsilon$ values are bias towards either positive or negative values) may very likely lead to differing results. We leave this as an open question, one possibly experimentally motivated, to see the impact of physical gates that may tend to produce errors with a bias.

\subsection{Smaller Scale Algorithms}
%

Based on the findings from the previous section, it is clear that the influences from P$_1$($\epsilon$) versus P$_2$($\epsilon$) are negligible towards the overall success of the algorithms. To confirm this fact, both the Grover and QFT circuits were tested as well, showing similar results. Consequently, we will choose to have all of the remaining results from this point forward be generated using only P$_1$($\epsilon$). The choice for using P$_1$($\epsilon$) is motivated by simplicity reasons, electing to work with a single Gaussian model versus a double.

\begin{figure}[h] 
\centering
\includegraphics[scale=0.4]{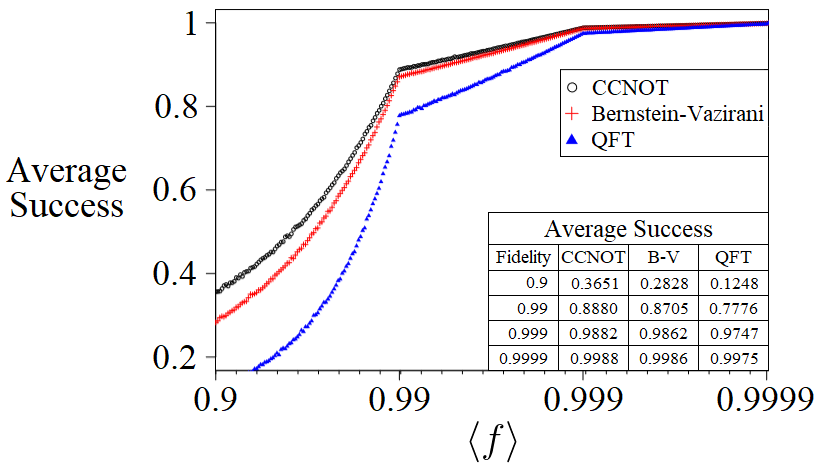}
\caption{Plotted are the average values of success for the CCNOT (black circle), Bernstein-Vazirani ( red +), and QFT(blue triangle) circuits. Each data point reflects the average success of the algorithms, generated from 10000 simulations per $\langle f \rangle$ value.}
\label{Fig:SmallFid}
\end{figure}

Having settled on P$_1$($\epsilon$) as the probability distribution for our simulations, let us now discuss the impact of these coherent error gates on the algorithms outlined in figures \ref{Fig:BVCircuit} - \ref{Fig:GroverCircuit}. We shall start by presenting the results for the Berstein-Vazirani, QFT, and CCNOT circuits, shown in figure \ref{Fig:SmallFid}. While the Bernstein-Vazirani algorithm is perhaps of little practical importance, the same cannot be said about the QFT and CCNOT circuits. Several of the quantum algorithms currently thought to be contenders for true quantum supremacy \cite{shor,qpe,vqe,qaoa} rely critically on quantum subroutines which require QFT and CCNOT. Thus, benchmarking their gate fidelity dependence is an important step towards realizing grander quantum algorithms.

For completeness, we must specify the way in which the simulations determine the success of each algorithm (which includes the results shown in figure \ref{Fig:P1vP2}). Starting with Bernstein-Vazirani, all qubits are initialized in the $|0\rangle$ state, and the success of the algorithm is based on the probability of measuring the state $|1010\rangle$. Specifically, let $|\Psi \rangle_f$ be the final state at the end of the circuit, which has absorbed all of the errors from the imperfect gate operations. Then, the success of the algorithm is the quantity $|\langle1010|\Psi\rangle_f|^2$.

For the CCNOT circuit, we define the measure of success as the quantity $|\langle 1100|\langle100|\Psi\rangle_f|^2$, where the control qubits are Q$_1$ and Q$_2$, and the target is a$_1$ (this state follows the structure $|Q_1 Q_2 Q_3 Q_4 \rangle|a_1 a_2 a_3 \rangle$, see figure \ref{Fig:QubitGeometry}). To produce this desired final state, we initialize qubits Q$_1$ and Q$_2$ in the state $|1\rangle$, and all other qubits in $|0\rangle$. These initialized qubits are done so perfectly in our simulation, ensuring that the only sources of error come from the gates outlined in $\ref{Fig:CCNOTCircuit}$.

For the QFT circuit, the measure of success is slightly different from the previous two. Because the QFT is often used in larger algorithms for the way in which it uniquely handles phases on each state, we have chosen to include phase into our model the success of our QFT simulations. To do this, we initialize the computational qubits in a specific superposition state $|\Psi \rangle_i$, which has a desired output state that contains no repeating phases:

\begin{eqnarray}
|\Psi\rangle_i &=& \frac{1}{2} \Big{(} |0011\rangle + |0111\rangle + |1011\rangle + |1111\rangle \Big{)} \nonumber \\
|\Phi \rangle &=& QFT |\Psi \rangle_i \nonumber \\
&=& \frac{1}{2} \Big{(} |0000\rangle -i |1000\rangle - |0010\rangle +i |0001\rangle \Big{)} \label{Eqn:QFTfinalstate}
\end{eqnarray}

The state shown in equation \ref{Eqn:QFTfinalstate} is used as our metric of success for the QFT circuit, $|\langle \Phi |\Psi \rangle_f|^2$. The initialization of $|\Psi \rangle_i$ is done using perfect gates, isolating the QFT circuit as shown in figure $\ref{Fig:QFTCircuit}$ as the only source of error. Additionally, because the QFT circuit does not end with a measurement, and in principle may be followed by further quantum operations, we also impose a strict condition on the ancilla qubits. Specifically, because of the way in which the quantum circuit is designed, we only consider final states where the ancilla qubits are returned to the state $|00\rangle$.

In all three circuits, the quantities of interest are represented by inner products squared, which can be observed experimentally (with the exception of the phases from QFT). However, because we are simulating these quantum systems classically, we have the advantage of being able to observe wavefunctions and amplitudes directly. Consequently, we can use the amplitudes of the desired final states to directly calculate the average probabilities of success, rather than simulating measurements. Figure $\ref{Fig:SmallFid}$ shows the results of our simulations, showing the average success rates for each circuit.

\begin{figure}[h] 
\centering
\includegraphics[scale=0.32]{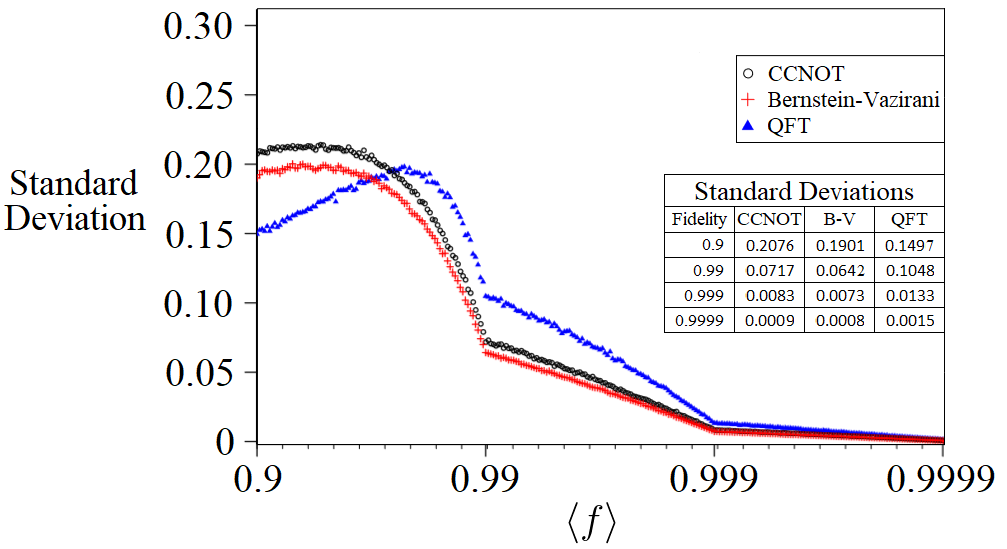}
\caption{Plotted are the standard deviations for the CCNOT (black circle), Bernstein-Vazirani (red +), and QFT (blue triangle) circuits. High standard deviations are a strong indicator that an algorithm's success is unreliable, prone to wildy differing results from run to run. Each data point corresponds to the same data used to generate the plots in figure \ref{Fig:SmallFid}.}
\label{Fig:SmallSigma}
\end{figure}

The algorithm results shown in figure \ref{Fig:SmallFid} were chosen due to their similarity in fidelity dependence. In particular, all three circuits show the largest increase in success in the region $0.9 \geq \langle f \rangle \geq 0.99$, becoming dependably successful by 0.999. In terms of current NISQ hardware, 99.9\% fidelity is certainly within the realm of feasibility, with perhaps the exception of the $\textbf{CNOT}$ gate. However, while the averages shown above may look promising, they do not tell the whole story. Figure \ref{Fig:SmallSigma} shows the standard deviations accompanying the results in \ref{Fig:SmallFid}, revealing that individual runs of these noisy circuits can vary drastically. Even for $\langle f \rangle$ values between 0.99 and 0.999, our simulations showed frequent individual trials with successes below 50\%, despite the averages being 85 - 98 \%.

When comparing the results in figure $\ref{Fig:SmallSigma}$ to $\ref{Fig:SmallFid}$, we can see that the CCNOT and Bernstein-Vazirani circuits have nearly mirror results. Intuitively, one might expect the CCNOT circuit to have smaller standard deviations due to its higher average success, but the data in figure $\ref{Fig:SmallFid}$ is actually revealing a critical feature.

Because the Bernstein-Vazirani algorithm uses Hadamard gates on five out of the seven qubits, nearly all of the algorithm's amplitude is spread evenly in superposition, only collapsing down to the $|1010\rangle$ state at the very end. During this superposition, the effects of the noisy gates appear to be distributed more evenly, leading to a smaller variance in final amplitudes. Conversely, because the CCNOT circuit deals with just three qubits, only one of which is in a superposition, the effects of the noisty gates tend to be more pronounced.

Lastly, the results of the QFT circuit seem to follow trends distinct from the other two, largely responsible by the increased size of the algorithm. The data from figure $\ref{Fig:SmallFid}$ clearly shows that the increased number of gate operations impedes the algorithm's success. Simultaneously, the notably higher standard deviations indicates that the circuit's complexity leads to consistently varying final states. The exception to this being the region where $\langle f \rangle < 0.99$, where we can attribute the small standard deviations to the algorithm's overall low average success.

\subsection{Larger Scale Algorithms}
%

Let us now turn our attention to the Grover Algorithm, which is considerably longer than the previous circuits. In the coming results, we will be examining the success of the Grover Algorithm at the point of each iteration. Much like the Bernstein-Vazirani and CCNOT circuits, the metric for success will be in the probability of measuring a single desired state, which by the design of the circuit will be the state $|0101\rangle$.

The only difference between the success metric here and the ones previously studied is that the theoretical desired final state does not have a probability of $1$. In particular, the theoretical probabilities of measuring the desired state are roughly $0.473$, $0.908$, and $0.961$ after one, two, and three Grover iterations respectively. As a result, we must adjust the success metric accordingly:

\begin{eqnarray}
\textrm{Grover Iterations} \hspace{1.2cm} \textrm{Success Metric} \nonumber \\
1 \hspace{2.4cm} \frac{1}{.473} | \langle 0101| \Psi \rangle_f |^2 \nonumber \\
2 \hspace{2.4cm} \frac{1}{.908} | \langle 0101| \Psi \rangle_f |^2 \nonumber \\
3 \hspace{2.4cm} \frac{1}{.961} | \langle 0101| \Psi \rangle_f |^2 \label{Eqn:Gsuccess}
\end{eqnarray}

\begin{figure}[h] 
\centering
\includegraphics[scale=0.4]{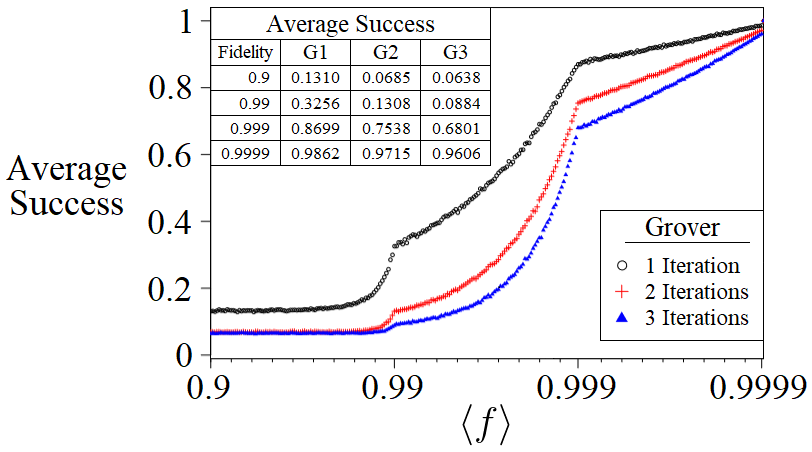}
\caption{Plotted are the average values of success for the Grover algorithm after one (black circle), two (red +), and three (blue triangle) iterations. Each data point reflects the average success of the algorithms for a particular $\langle f \rangle$ value, generated from 2000, 1500, and 1000 simulations per value for one, two, and three iterations respectively.}
\label{Fig:Gplots}
\end{figure}

Using the adjusted success metrics shown in equation \ref{Eqn:Gsuccess} (in the actual simulations we compare using values of higher decimal accuracy), we can track the success of the Grover Algorithm after each iteration. Plotted in figure $\ref{Fig:Gplots}$ are the results found from our simulations.

In contrast to the plots for the smaller algorithms, the success of the Grover Algorithm is noticeably worse. This result is perhaps unsurprising, considering that the Grover Algorithm is several times larger in gate count, even containing several CCNOT gates within it. If we compare the success of this algorithm at the average fidelity point 99.9\%, we can see that even optimistic quantum computing hardware would produce unreliable results.

\begin{figure}[h] 
\centering
\includegraphics[scale=0.38]{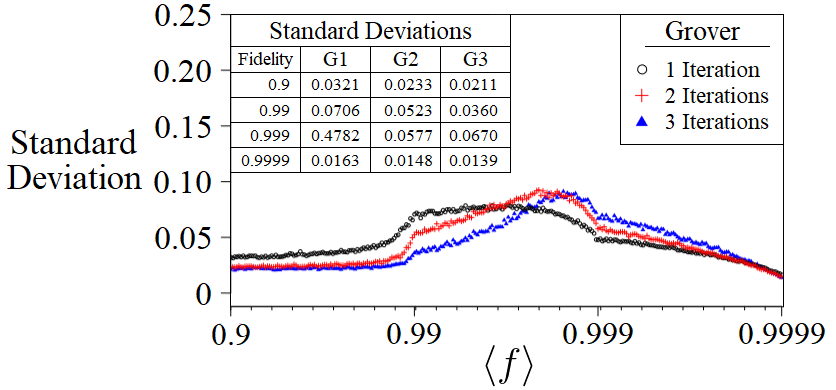}
\caption{Plotted are the standard deviations for the three Grover iterations. Each data point corresponds to the same data used to generate the plots in figure \ref{Fig:Gplots}, but without the factors for the success metric. Thus, the standard deviation values shown here represent the variance in amplitude found on the $|0101\rangle$ state at the end of each iteration.}
\label{Fig:Gsigmas}
\end{figure}

When we compare the trends shown in figures \ref{Fig:SmallFid} and \ref{Fig:Gplots}, one way to look at the respective successes is in terms of the order of magnitude where we see the largest growth. While the smaller algorithms see the largest gain in success in the average fidelity region [0.9 - 0.99], the Grover Algorithm achieves similar growth in the [0.99 - 0.999] region. Thus, we can say that the success of the Grover Algorithm requires an additional order of magnitude in gate fidelity. To confirm this result once more, figure $\ref{Fig:Gsigmas}$ shows the accompanying standard deviations to the Grover plots.

The trends found in figure $\ref{Fig:Gsigmas}$ are in agreement with the trends shown in $\ref{Fig:Gplots}$. The data shown in this figure represents the standard deviations for the unadjusted probabilities of measuring the $|0101\rangle$ state (the same quantities in equation \ref{Eqn:Gsuccess} but without the prefactors). We can see that each Grover iteration undergoes the same peak in standard deviation in the region where $ 0.99 < \langle f \rangle < 0.999 $. However, unlike the trends seen with the smaller algorithms, the Grover Algorithm doesn't find reliable results until the gate fidelities are beyond 99.99\%.

%
%
\section{Modeling Decoherence Errors}
%

It is important to remember that imperfect gates are just one source of error that plague NISQ computers. A second major source can be categorized as decoherent errors. In contrast to the coherent errors studied up to this point, these are errors which occur spontaneously and often times cannot be described by unitary operators. There are several well documented models for the source of these decoherence errors \cite{T1_T2,T2_1,T2_2}, but our interest in this study will be primarily energy relaxation and partial wavefunction collapses, and their impact on algorithm success.

When a quantum system experiences a decoherence error, what we will mean is that a qubit (or multiple qubits) has undergone a collapse in some way, a process which irreversibly disturbs the system.  The first of two such errors which we will simulate is known as energy relaxation, parameterized by the metric $T_1$, and represents a qubit in the $|1\rangle$ state (excited state) collapsing to $|0\rangle$ (ground state).  While the name ``$T_1$'' is sometimes interchangeably used to describe the error it represents, strictly speaking $T_1$ is a length of time (referred to as coherence time). Mathematically, $T_1$ represents the amount of time a qubit can be expected to probabilistically hold onto its excited state, governed by the exponential decay probability function given in equation \ref{Eqn:P(t)}:

\begin{equation}
\textrm{Probability of \textit{no} error:} \hspace{.5cm} \textrm{P}(\Delta t) = \mathLarge{ e^{\frac{-\Delta t}{T_j}} }
\label{Eqn:P(t)}
\end{equation}

According to this probabilistic function, the value $T_1$ corresponds to the length of time where one expects that a given qubit $\textit{has}$ collapsed with a probability of $1 - 1/e$ (roughly 63\%). The equation tells us that the chances of such an energy relaxation occuring decrease with either shorter algorithm times ($\Delta t$) or longer $T_1$ coherence times. While better decay rates will certainly be realized as technology continues to improve, the same cannot be said for algorithm times. Algorithms can always be optimized to try and minimize $\Delta t$, but in principle we should expect that future advanced algorithms will inherently require larger $\Delta t$'s.

%
\subsection{Simulating $T_1$ and $T_1^{*}$}
%

In addition to energy relaxation, in this study we will also simulate a second type of decoherent error, akin to a partial measurement collapse.  The motivation for including such events into our simulations is to test a wider range of possibilities for a noisy quantum system to fail, and the corresponding impact on algorithm success.  Analagous to the energy relaxation error, we model the probability of a partial collapse occurring on a given qubit by a single parameter, which we will denote $T_1^{*}$, obeying the same probability decay as in equation \ref{Eqn:P(t)}.  Consider the example below, which shows how a quantum system would collapse under $T_1$ and $T_1^{*}$ errors in our simulations:

\begin{eqnarray}
| \Psi \rangle_i = \alpha |010\rangle + \beta |110\rangle + \gamma |011\rangle \\
\nonumber \\
T_1: \hspace{.3cm} \textrm{qubit}_2 \longrightarrow |0\rangle \hspace{.9cm} \nonumber \\
| \Psi \rangle_f = \alpha |000\rangle + \beta |100\rangle + \gamma |001\rangle \label{Eqn:T1} \\
\nonumber \\
T_1^*: \hspace{.3cm} \textrm{qubit}_1 \longrightarrow |0\rangle \hspace{.9cm} \nonumber \\
| \Psi \rangle_f = \frac{1}{\sqrt{ \alpha^2 + \gamma^2 }} \Big{(} \alpha |010\rangle + \gamma |011\rangle \Big{)} \label{Eqn:T1*}
\end{eqnarray}

Beginning with the $T_1$ error, equation \ref{Eqn:T1} shows the result of qubit 2 ($|\hspace{.02cm}q_1 q_2 q_3 \hspace{.02cm}\rangle$) collapsing from the $|1\rangle$ state down to $|0\rangle$. In our model, this type of error is only applicable when a qubit is solely in the $|1\rangle$ state (no superposition), as can be seen in the state $| \Psi \rangle_i$.  The motivation for this choice is to instead elect for all qubit collapses from superposition states to be handled by $T_1^*$, whereby the probability of a qubit collapsing into either the $|0\rangle$ or $|1\rangle$ state is determined by amplitudes at the moment of the collapse.  When such an event does occur in our simulations, the resulting quantum state is then normalized based on the result, as shown in \ref{Eqn:T1*}.  Note that once a qubit has collapsed to the $|1\rangle$ state as a result of a $T_1^*$ error, it is then immediately subject to $T_1$ energy relaxation thereafter. 

\begin{figure}[h] 
\centering
\includegraphics[scale=0.5]{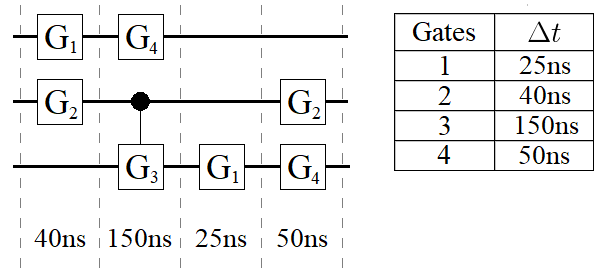}
\caption{Example circuit showing the resulting times for each moment, based on different pairings of gates. The operation with the longest gate time in a given moment determines the total time for that moment. }
\label{Fig:DecoherenceTimes}
\end{figure}

In determining when and where decoherence errors occur, our simulations work through each algorithm in ``moments'' to determine the $\Delta t$ for equation $\ref{Eqn:P(t)}$. A moment is defined as any grouping of gates that can happen in parallel, represented pictorially by gates in a vertical stack in figures \ref{Fig:BVCircuit} - \ref{Fig:GroverCircuit} (with the exception of a few CCNOT gates in the Grover Circuit that are drawn in separate moments for display purposes). Because these gates would be physically occurring at the same time, our simulations treat them in the same way. Specifically, the $\Delta t$ for all of the qubits in a given moment is determined by the longest gate in the moment. Figure \ref{Fig:DecoherenceTimes} shows an example of this using gates with varying $\Delta t$'s.

For our simulations, we implement occurrences of decoherence errors after all of the gates in a given moment. Since it is unclear what the model for a decoherence error $\textit{during}$ a gate implementation would be, we will elect to let gate operations happen independent of decoherence errors. Thus, after a grouping of gates have been applied in parallel, our simulation works through each qubit and randomly samples $P(\Delta t)$ based on the $\Delta t$ for that moment. Note that splitting up the $\Delta t$ times as shown in figure \ref{Fig:DecoherenceTimes} is mathematically correct because of the exponential nature of these spontaneous errors \ref{Eqn:P(t)}. As shown in equation \ref{Eqn:Deltat}, we are guaranteed that sequentially sampling from this distribution is equivalent to sampling the same total time interval:

\begin{equation}
\prod_i e^{-\Delta t_i / T_j} = e^{-\Delta t_{\textrm{total}} / T_j} \hspace{.8cm} \Delta t_{\textrm{total}} = \sum_i \Delta t_i
\label{Eqn:Deltat}
\end{equation}

\begin{figure}[h] 
\centering
\includegraphics[scale=0.5]{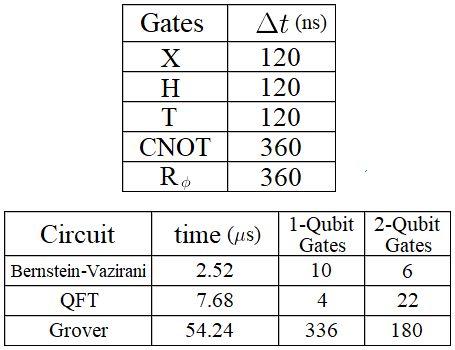}
\caption{(top) Gate times for the 1 and 2-qubit gates used in the simulations. (bottom) Breakdown of each algorithm's total time as well as the number of 1 and 2-qubit gates. In determining the total length of time for a circuit, one must consider both the total number of moments as well as types of gate in each moment.}
\label{Fig:GateTimes}
\end{figure}

Using equation \ref{Eqn:Deltat} along with the circuit diagrams $\ref{Fig:BVCircuit}$ - $\ref{Fig:GroverCircuit}$, we can calculate the total times required for each algorithm. Figure \ref{Fig:GateTimes} shows the times that we have chosen for the 1 and 2-qubit gates, as well as the resulting total times for each algorithm. These times are based on average results found from reports for 1 and 2-qubit gates on superconducting qubits.

%
%
\section{Decoherence Analysis}
%

We will now present the results of our decoherence simulations here, focusing on noteworthy trends in the data. In all of the coming simulation results, the only sources of error for each quantum system are $T_1$ and $T_1^*$ collapses. All of the gate operations for this section assume perfect fidelity as to isolate the impact of the decoherence errors. In addition, for all reported $T_1^*$ times we set the value of $T_1 = 2 T_1^*$ when otherwise unspecified, a choice made for simplicity reasons in our simulations. Lastly, we apply the same values for $T_1$ and $T_1^*$ to each qubit, assuming all qubits are of equal quality.

%
\subsection{Algorithm Success}
%

\begin{figure}[h] 
\centering
\includegraphics[scale=0.39]{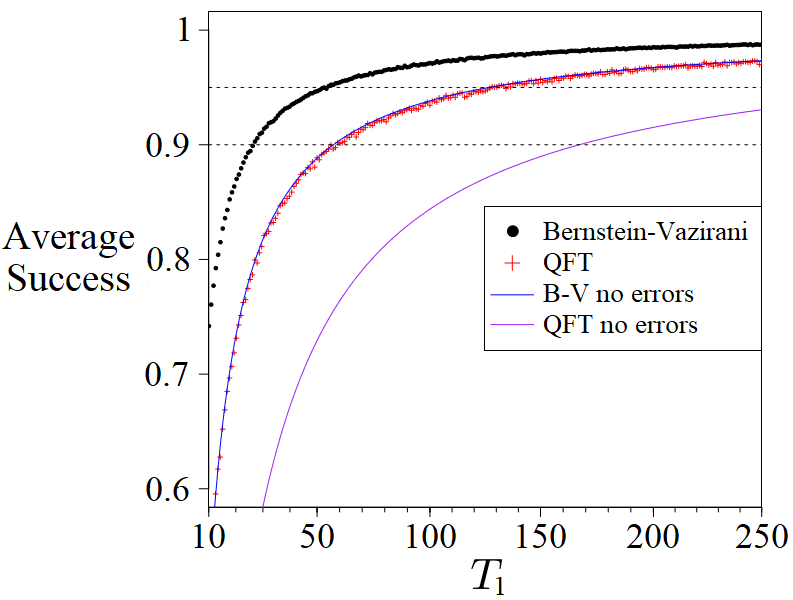}
\caption{The average success rates of the Bernstein-Vazirani (black circle) and QFT (red +) as a function of coherence time $T_1$, with $T_1 = 2T_1^*$. Alongside the data are plots for P($\Delta t$), using the total $\Delta t$ for each circuit from $\ref{Tbl:T2_Times}$. These curves represent the probability of having zero decoherence errors as a function of increasing coherence times.}
\label{Fig:T2_BV_QFT}
\end{figure}

Analogous to the results shown in figures $\ref{Fig:SmallFid}$ and $\ref{Fig:Gplots}$, our first quantity of interest will be the relation between coherence times and algorithm success. Because these decoherence errors are spontaneous, one can expect to have entire runs with no errors, and some with multiple. Thus, our first result will be to run each algorithm numerous times and derive trends in average algorithm success.  Plotted in figure $\ref{Fig:T2_BV_QFT}$ are the average success rates for the Bernstein-Vazirani and QFT circuits as a function of $T_1$ (and correspondingly $T_1^*$ = $T_1/2$). Just like the fidelity simulations from earlier, the metric for success of each algorithm is defined as $|\langle \Phi| \Psi \rangle _f |^2$. As before, $|\Phi \rangle$ is the desired final state for each algorithm and $| \Psi \rangle _f$ is the final state of the system prone to errors.

As one might expect, the simulations show that the Bernstein-Vazirani circuit is less prone to coherence errors as a result of having a shorter total circuit time. However, unlike the data from the fidelities errors, we do not see the two plots converging to $1$ quite as quickly. Despite being the smallest two algorithms, their success rates only reach $98.7$\% and $97.3$\% for the maximum studied value of $T_1 = 250 \mu$s.

\begin{figure}[h] 
\centering
\includegraphics[scale=0.39]{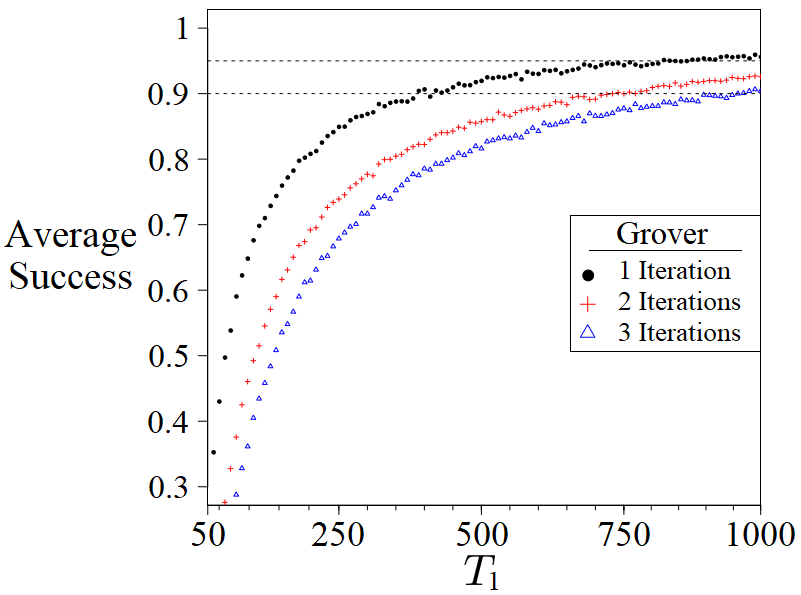}
\caption{The average success of the Grover Algorithm at the one (black circle), two (red +), and three (blue triangle) iteration points as a function of coherence times $T_1$. Marked along each plot are the success benchmarks for 90\% and 95\%.}
\label{Fig:T2_Grover}
\end{figure}

Also plotted in figure $\ref{Fig:T2_BV_QFT}$ are curves which show the probability of no error occurring for each circuit as a function of $T_1$. The reason these additional plots are of interest is because they represent the scenario in which a single decoherence error results in a $0\%$ success probability for an algorithm. So then, the large discrepancy between these curves and the data points is indicative that these decoherence errors do not completely kill an algorithm. In the next section we will explore this topic in further detail, but first we will continue our preliminary analysis by looking at data for the Grover circuit(s), shown in figure \ref{Fig:T2_Grover}.

By comparison to the results from the Bernstein-Vazirani and QFT circuits, the data in figure $\ref{Fig:T2_Grover}$ shows a significantly worse trend. Whereas the two smaller circuits reach the $90\%$ success mark for $T_1$ coherence times as short as $30 \mu$s and $61 \mu$s, the Grover iterations do not achieve such success until $425\mu$s, $745\mu$s, and $975 \mu$s. Just like the case of the coherent amplitude errors, we find that the increased size of the Grover circuit causes the algorithm to find success roughly an entire order of magnitude later than the smaller algorithms.

%
\subsection{Total Decoherence Time}
%

\begin{figure}[h] 
\centering
\includegraphics[scale=0.36]{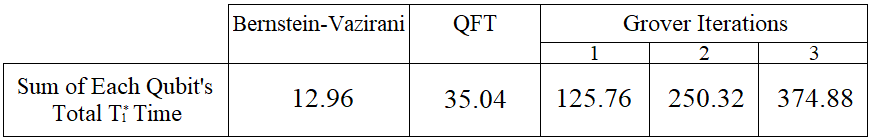}
\caption{The sum of the total amount of time in each circuit for which a qubit is prone to a $T_1^*$ decoherence error. Instances where $T_1$ collapses may occur were purposely excluded in order to show the total amount of time in which each algorithm must sustain superpositions.}
\label{Tbl:T2_Times}
\end{figure}

To better understand the plots shown in figures $\ref{Fig:T2_BV_QFT}$ and $\ref{Fig:T2_Grover}$, it is helpful to not only consider the total time for the entire circuit, but also the total individual times for which each qubit may undergo a decoherence collapse. For example, table $\ref{Fig:GateTimes}$ shows that the difference in total time between the Bernstein-Vazirani and QFT circuits is roughly 5$\mu$s, but when we sum up the total amount of time in the circuits for which each individual qubit must sustain a superposition, we find the difference between the two circuits to be nearly 22$\mu$s. Table $\ref{Tbl:T2_Times}$ shows the total times for which each circuit must endure spontaneous superposition collapse errors.

Based on the way in which we have chosen to model spontaneous decoherence errors, the numbers shown in figure $\ref{Tbl:T2_Times}$ represent the primary governing factor for the likeliness of an error in each algorithm. That is to say, substituting the times from this table into equation $\ref{Eqn:P(t)}$ and plotting as a function of $T_1^*$ will reveal curves that are nearly identical to those plotted in figure $\ref{Fig:T2_BV_QFT}$ (the curves for no error). By comparison, our simulations found results which are also very close in shape to exponential curves, despite being averages that incorporate many different trials spanning various combinations of errors.

%
\subsection{Impact of Single Decoherence Errors}
%

While the graphs from the previous section are good indicators into the relation between spontaneous collapses and algorithm success, here we will delve a bit deeper into the exact nature of what these types of errors may do to the quantum systems. Often times a single decoherence error is assumed to be the death of an algorithm, but this is not necessarily always the case. Certain algorithms could in principle be designed such that a decoherence error on particular qubits has a tolerable impact on the overall success of the algorithm.

\begin{figure}[h] 
\centering
\includegraphics[scale=0.26]{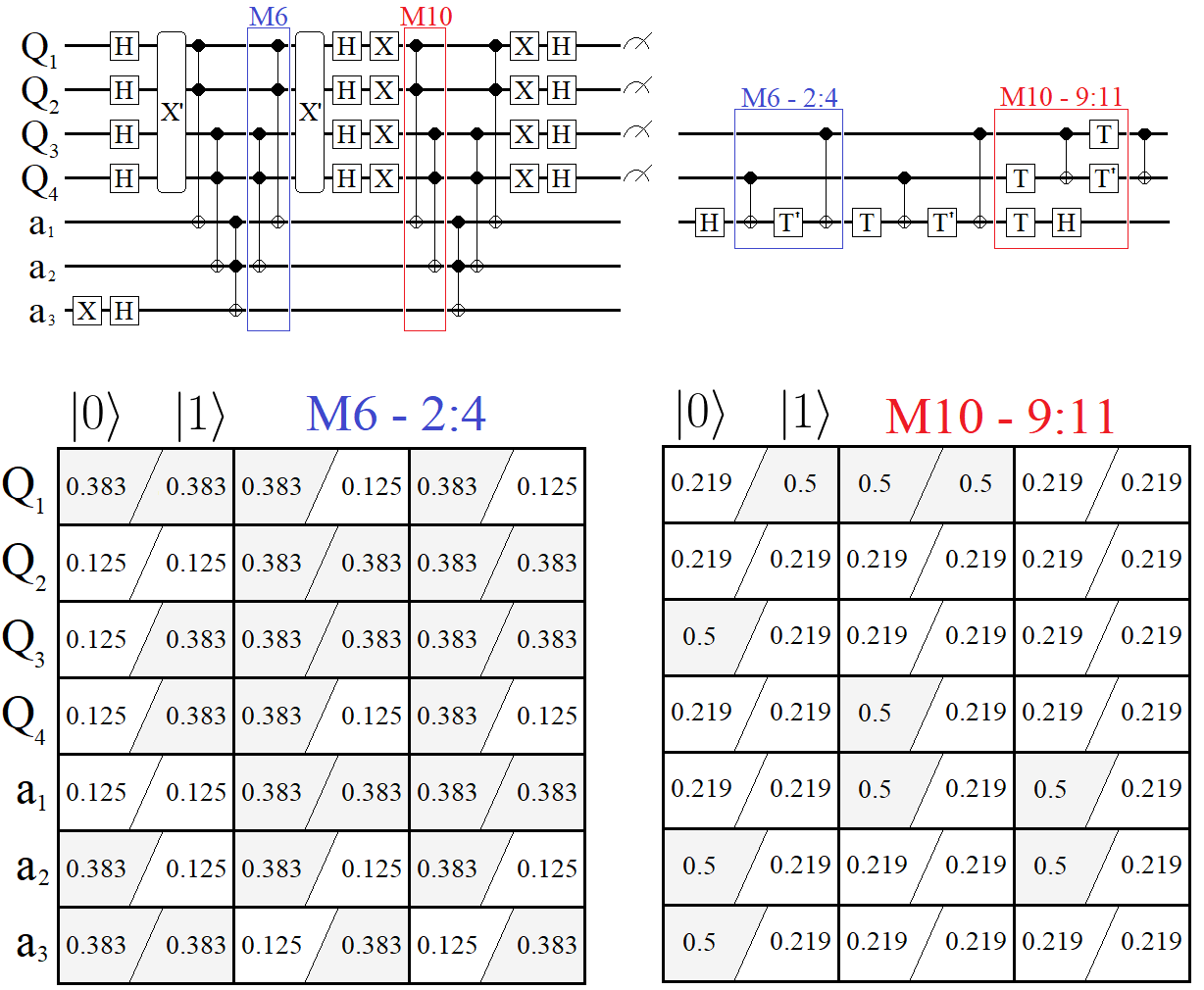}
\caption{Success rates for the Grover Algorithm (1 Iteration) for the situation where exactly one partial measurement collapse has occurred, chosen for two distinct moments in the circuit (here we see examples of which CCNOT gates happen in parallel). The numbers in each box represent the overall success of the algorithm, defined as in equation $\ref{Eqn:Gsuccess}$. The left value in each box is for the case where the qubit collapses to the $|0\rangle$ state, and similarly the $|1\rangle$ state for the right value.}
\label{Fig:G_1E}
\end{figure}

For example, consider the way in which we use ancilla qubits in the circuits for this study, often only serving a temporary purpose in the form of CNOT gates. After successfully delivering the effect of a CNOT gate between two distant computational qubits, a decoherence error on them may have little to no impact on the overall algorithm. To demonstrate that not all decoherence errors are equal, figure $\ref{Fig:G_1E}$ shows how the impact of a $T_1^*$ collapses can vary depending on which qubit or moment it occurs on.

The tables in $\ref{Fig:G_1E}$ show that a single decoherence error is not necessarily fatal to an algorithm, depending on the location of the error and the resulting value of the collapse. This is shown both within the two tables in the figure, as well as between them. For the case of a single error occurring within the overall 6$^{th}$ moment (left table), we see the recurring values 0.125 and 0.383 show up depending on which qubits undergo the collapse to the $|0\rangle$ or $|1\rangle$ state. Even within a single qubit, there are instances where a collapsed value of $|0\rangle$ may be more tolerable, while collapsing to $|1\rangle$ is preferable on the very next moment.

Similarly, if we compare the values between the tables, we find that errors occurring during a later portion of the algorithm result in completely different success rates. If we look at the values in the right table, corresponding to the overall 10$^{th}$ moment (and again the sub moments within the CCNOT circuit), we now find that the success rates of the algorithm vary between 0.219 and 0.5 depending on the location and value of the collapse. These numbers tells us two interesting things: 1) The overall success of this particular algorithm is more resilient if a single decoherence error were to occur in this later moment. 2) There are select instances where a decoherence error actually results in a $\textit{better}$ final state.

To understand this second point, recall that 1 iteration of the Grover Algorithm results in a probability of measuring the desired state of about 47.3\% (equation $\ref{Eqn:Gsuccess}$). This probability comes from a final state where the $|0101\rangle$ state is most probable, and all other $15$ states in the system share the remaining probability. So then, if now we imagine that a spontaneous partial collapse were to happen on a qubit at the end of the circuit, resulting in the desired final state for that qubit, this would in turn remove a portion of the non-desired states from the system and consequently boost the overall probability of the desired state.  Decoherence errors collapsing in favorable ways is certainly a rarity, and in general only applicable to certain algorithms. For example, there is no single collapse which can boost an algorithm with a desired final state probability of 1 such as Bernstein-Vazirani. Nevertheless, these rare collapses in the Grover Algorithm further the claim that not all decoherence errors are fatal to an algorithm.

%
\subsection{Success By Error Count}
%

\begin{figure}[h] 
\centering
\includegraphics[scale=0.38]{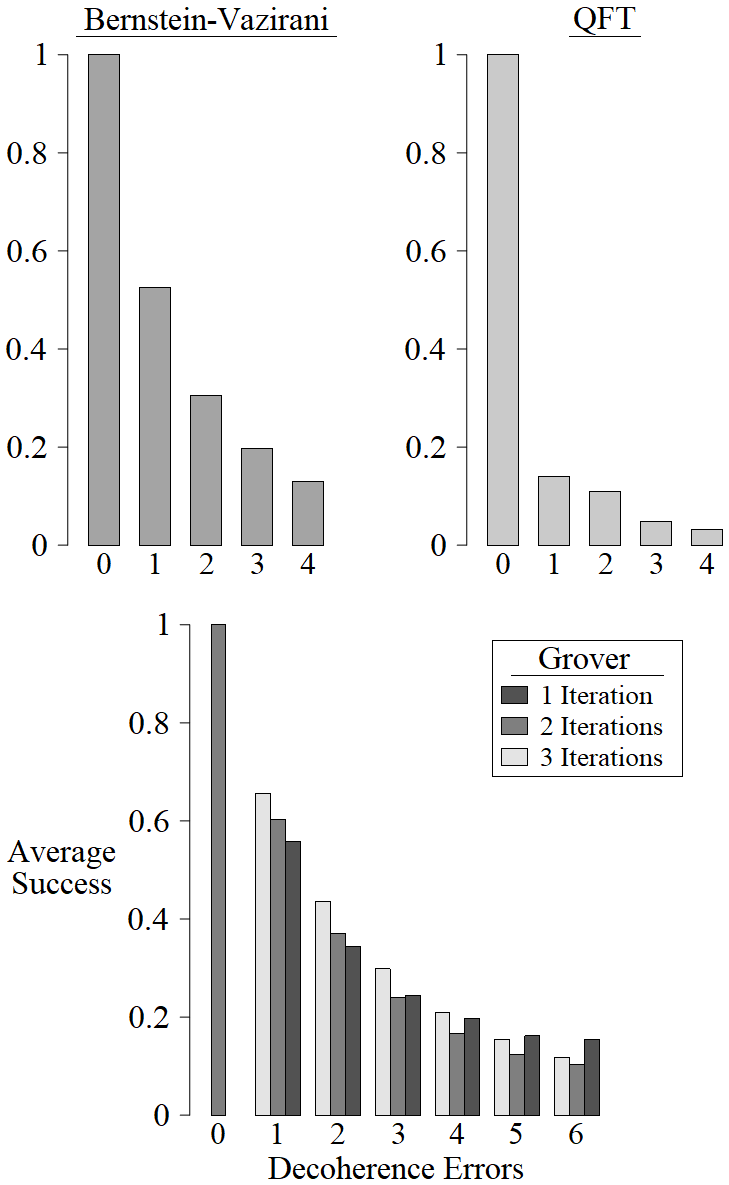}
\caption{Plotted are the average success rates for the Bernstein-Vazirani (top left), QFT (top right), and Grover (bottom) circuits as a function of total number of decoherence errors. For the Grover Algorithm, each iteration is plotted as its own color, highlighting the resilience of the algorithm for the various lengths. All three plots show that the largest drop in success occurs from the first error.}
\label{Fig:T2_Count}
\end{figure}

Having just seen some examples of algorithm success for cases of exactly one collapsing error, we will now turn our attention to the impact of numerous decoherence errors. Figure $\ref{Fig:T2_Count}$ shows the average success of each algorithm as a function of the total number of decoherence errors. These plots were generated from the same results used in average date trends in figures $\ref{Fig:T2_BV_QFT}$ and $\ref{Fig:T2_Grover}$, but now separated by instances of various error counts.

The bar plots in figure $\ref{Fig:T2_Count}$ show the rate at which numerous decoherence errors impede the overall success of each algorithm. For all three of the circuits studied, it is clear that the biggest cost in algorithm success comes from the first decoherent collapse a quantum system experiences, after which each successive error has a diminishing effect. This turns out to be especially true for the QFT circuit, which can be seen as the least resilient algorithm to a single decoherence error. Ultimately, given enough decoherence errors in a single run, we can see that the quantum systems reach a point where the effect of each algorithm is completely washed out and we are left with probabilities nearing an equal distribution of all states.

If we now compare the results in figure $\ref{Fig:T2_Count}$ with the additional plots in figure $\ref{Fig:T2_BV_QFT}$, we can see why the average success rates are higher than those of the zero error curves. Specifically, we can think of the data from $\ref{Fig:T2_BV_QFT}$ as showing the combined average of each success rate from $\ref{Fig:T2_Count}$, multiplied by the weight of that many errors occuring. As we increase the coherence times of $T_1$ and $T_1^*$, we not only increase the probability of getting a run with zero errors, but also decrease the occurrence of multiple errors and correspondingly the lower success rates contributing to the overall average.

%
%
\section{Combining Error Models}
%

We have now seen the effects of the two models for error studied in this paper: coherent amplitude errors and collapsing decoherence errors. These two error models were derived to solely incorporate two of the most commonly reported values for benchmarking quantum computers: $\langle f \rangle$ and $T_1$. In this final section, we will combine both of these error models and study their joint impact on algorithm success.

Based on the results from studying each error model in isolation, we have chosen to study their combined effect in a way which assumes a continuous improvement in both parameters. Specifically, each data point in the coming figure represents a consistent percentile improvement in both average fidelity and coherence times from the previous point. The values for $\langle f \rangle$ and $T_1$ will obey the following trends:

\begin{eqnarray}
\textrm{Initial Values:} \hspace{.5cm} \langle f \rangle_i &=& 0.99 \hspace{.5cm} T_{1i} = 20\mu s \nonumber \\
\langle f \rangle_k &=& (0.9) \hspace{.05cm} \langle f \rangle_{k-1} + 0.1 \\
T_{1 k} &=& (1.05) \hspace{.05cm} T_{1k-1}
\end{eqnarray}

As before, we set the value of $T_1^*$ to be half that of $T_1$ for all points. The motivation for combining the errors in this way is to simulate what one might expect from technological improvements on a continual basis. These chosen values then represent the scenario in which gate fidelities and coherence times improve at rates of $10\%$ and $5\%$ respectively, which is not unreasonable given past trends in technological improvements.

\begin{figure}[h] 
\centering
\includegraphics[scale=0.4]{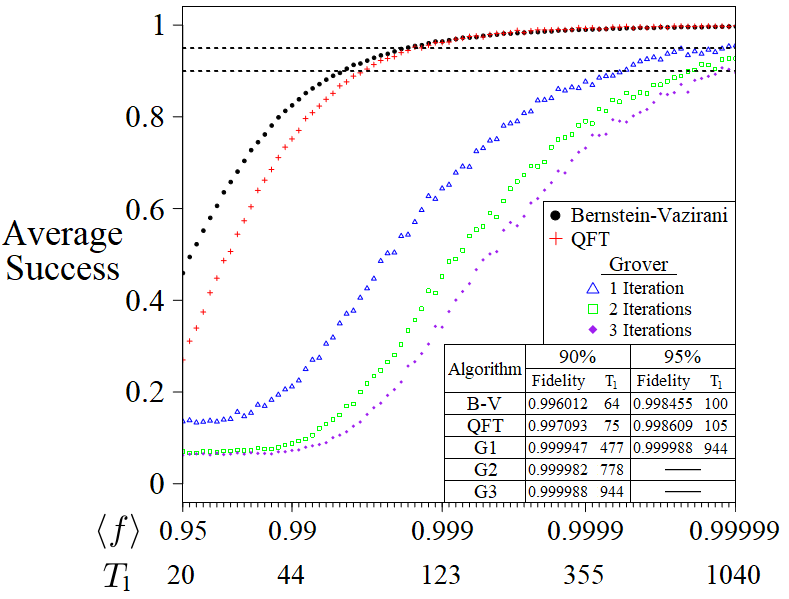}
\caption{Average success rate as a function of both sources of error. Each data point in the plot represents a 10\% increase in average fidelity and 5\% increase in coherence times from the previous point. The accompanying table shows the points at which each algorithm crosses the 90\% and 95\% average success threshold, and the corresponding $\langle f \rangle$ and $T_1$ values.}
\label{Fig:Both_Errors}
\end{figure}

Figure $\ref{Fig:Both_Errors}$ shows that the result of incorporating both error models results in noticeably worse results. Interestingly, if we compare the marked thresholds of success to those in figures $\ref{Fig:T2_BV_QFT}$ and $\ref{Fig:T2_Grover}$, we find that the combined error trends are closer to those of the isolated decoherence errors. This result suggests that between the two types of errors, the decoherence errors seem to outweigh the coherent amplitude errors in impeding algorithm success.

If we now consider where current NISQ devices would fall on the x-axis shown in figure $\ref{Fig:Both_Errors}$, leading quantum computing efforts could be categorized as somewhere in the region between (0.99,87) and (0.999,255). If we focus on the various successes found within this region, the results indicate promising results for smaller algorithms such as QFT. This in turn suggests that algorithms which can be composed of 20-30 gate operations or less may find reasonable success in the near future.

%
%
\section{Conclusion}
%

The results found from the various simulations in this paper explore the degree to which imperfect gate operations and decoherence errors are detrimental to algorithm success. For the case where the only source of error in the system is imperfect gates, it was found that the necessary fidelities in order to achieve average success rates of greater than 90\% ranged from $0.99 \geq \langle f \rangle \geq 0.999$ for the smaller algorithms, and upwards of 0.9999 for the Grover iterations. Similarly, in order to achieve the same levels of average success with only decoherence errors, the simulations found that the required coherence times were of the order $50 \mu s \geq T_1 \geq 500 \mu s$.

%
\subsection{Impact By Algorithm}
%

The results from the isolated error cases suggest that the smaller algorithms (Bernstein-Vazirani, CCNOT, and QFT) may find reasonable success on current NISQ devices. However, the results from the combined errors simulations showed that even these smaller circuits may be just barely on the cusp of feasible, requiring a combination of gate fidelities and coherence times around the order of $0.997$ and $80 \mu s$. When we compare these values to that of the latest state-of-the-art quantum devices, which promise $\langle f \rangle$ and $T_1$ values around $0.995$ and $50 - 100 \mu s$, it is difficult to imagine even the smaller algorithms achieving the 90\% average success percentile.

In regards to the Grover Algorithm, and its success as a function of iterations, the results from all three studies concluded that such a quantum circuit is beyond the reach of current technology. In particular, in order to run circuits with the same level of depth and gate count as those studied in this paper, our simulations show that the critical quality for improvement is $T_1$. This result was also found to be consistent for the smaller algorithms as well, which suggests that technological improvements in coherence times will likely result in the biggest jumps in success for near term devices. Conversely, the results from the isolated fidelity study suggest that NISQ devices may already optimistically be in the region where $\langle f \rangle$ can produce reliable results. That being said however, our results assumed that CNOT gates could perform on the same order of precision as single qubit gates, which has yet to be demonstrated experimentally.

%
\subsection{Benchmark Parameters, Models, and Future Work}
%

The two models for error studied in this paper can be interpreted as first-order approaches to understanding the impact of noise on quantum algorithm success. In particular, the average success rates found for the various algorithms are most indicative of circuit depth and gate count. Thus, the results shown in the figures throughout this study represent estimates to the orders of magnitude on $\langle f \rangle$ and $T_1$ one might require in order to expect reliable results.  However, being in the unique time in quantum computing we are, NISQ devices require much more detailed characterization then simply a handful of metrics in order to construct meaningful circuits. Specifically, because noise is such an unavoidable entity with these devices, any hopes of achieving near term quantum advantages will likely require algorithms that directly account for and minimize errors down to each individual qubit.

Going forward, there is room for several areas for improvement in the error models studied in this paper in order to yield results closer to what one might find on a physical device. The most notable improvement would be to vary the values of $\langle f \rangle$ and $T_1$ for the different gates and qubits (in principle, each qubit could have a complete list of $\langle f \rangle$ values for every gate operation), as well as the addition of the depolarizing noise model described by $T_2$ \cite{T2_1,T2_2}. For future research efforts with specific hardware parameters in mind, it would be interesting to see the accuracy of such advanced models as compared to physical results, where each qubit is customized to match the specifications of a real device.

\section*{Acknowledgement}

We gratefully acknowledge support from the National Research Council Associateship Programs and funding from OSD ARAP QSEP program. Additionally we would like to thank the STEM Outreach program, in collaboration with the Griffiss Institute of Technology, for supporting this research. Any opinions, findings, conclusions or recommendations expressed in this material are those of the author(s) and do not necessarily reflect the views of AFRL.

\pagebreak

\end{document}